\documentclass[12pt,a4paper]{article}
\usepackage{epsfig}
\usepackage{amssymb}

\title{Wrapped membranes, matrix string theory \\
and an infinite dimensional Lie algebra}
\author{
{\sc Shozo Uehara}\footnote{e-mail:
uehara@eken.phys.nagoya-u.ac.jp}~ and
{\sc Satoshi Yamada}\footnote{e-mail:
yamada@eken.phys.nagoya-u.ac.jp}\vspace{4mm}\\
{\sl Department of Physics, Nagoya University}\\
{\sl Chikusa-ku, Nagoya 464-8602, Japan}}
\date{}

\makeatletter

\@addtoreset{equation}{section}
\renewcommand{\thefigure}{\@arabic\c@figure}
\makeatother

\newcommand{\nn}{\nonumber\\}
\newcommand{\ptau}{\partial_\tau}

\newcommand{\ptheta}{\partial_\theta}
\newcommand{\mtheta}{\mathit{\Theta}}

\newcommand{\p}{\partial}

\newcommand{\DS}{\displaystyle}

\newcommand{\scom}[1]{[\,#1\,]_*}
\newcommand{\Lag}{\mathcal{L}}

\begin{document}
\maketitle
\vspace{-80mm}
\begin{flushright}
DPNU-04-02\\
hep-th/0402012
\end{flushright}
\vspace{57mm}

%%%%%%%%%%%%%%%%%%%%%%%%%%%%%%%%%
\begin{abstract}
%%%%%%%%%%%%%%%%%%%%%%%%%%%%%%%%%
We examine the algebraic structure of the matrix regularization for
the wrapped membrane on $R^{10}\times S^1$ in the light-cone gauge.
We give a concrete representation for the algebra and obtain the
matrix string theory having the boundary conditions for the matrix
variables corresponding to the wrapped membrane, which is referred to
neither Seiberg and Sen's arguments nor string dualities.
We also embed the configuration of the multi-wrapped membrane in
matrix string theory.
\end{abstract}
%%%%%%%%%%%%%%%%%%%%%%%%%%%%%%%%%%
\section{Introduction}
%%%%%%%%%%%%%%%%%%%%%%%%%%%%%%%%%%
It is believed that the supermembrane in eleven dimensions \cite{BST}
plays an important role to understand the fundamental degrees of
freedom in M-theory which is a unified description of various
superstring theories.
Actually, the matrix-regularized theory \cite{Hop,dWHN} of the
light-cone supermembrane, which is called Matrix theory, is
conjectured to describe light-cone quantized M-theory in the large-$N$
limit \cite{BFSS}.
Furthermore, even at finite $N$, Matrix theory is conjectured
to describe the $p^{+}=N/R$ sector of discrete light-cone quantized
(DLCQ) M-theory \cite{Sus}.\footnote{In this paper we use a convention
of the light-cone coordinates $x^\pm\equiv (x^{0}\pm x^{10})/\sqrt2$.
Furthermore, $x^-$ is compactified on $S^1$ with radius $R$ in DLCQ.}

Matrix string theory \cite{Mot,DVV} was proposed on the heels of
Matrix theory conjecture.
This theory is the 1+1-dimensional $U(N)$ super Yang-Mills theory and
it is conjectured to be a non-perturbative formulation of light-cone
quantized type-IIA superstring theory in the large-$N$ limit.
The theory is also conjectured to describe the $p^{+}=N/R$ sector of
DLCQ type-IIA superstring theory even at finite $N$ \cite{Sus}.
The proposal of matrix string theory is explained, on the basis
of Seiberg and Sen's arguments \cite{Sei,Sen},
by using the T- and S-dualities with the 9-11 flip of interchanging
the role of the 11th and 9th directions \cite{Mot,DVV}.

On the other hand, type-IIA superstring in ten dimensions can be
regarded as double-dimensional reduced supermembrane in eleven
dimensions \cite{DHIS}.\footnote{The double-dimensional reduction was
discussed classically in Ref.\cite{DHIS}. In quantum mechanically, it
is subtle whether such a reduction is realized or not
\cite{Rus,SY,UY}.}
Hence, it is natural to think that matrix string theory can be
regarded as the matrix-regularized theory for the wrapped
supermembrane on $R^{10}\times S^1$ in the light-cone gauge.
Actually, the correspondence between the wrapped supermembrane and
the matrix string was given in Ref.\cite{SY}.
Then, more systematic derivation of matrix string theory by the matrix
regularization of the wrapped supermembrane was presented \cite{Ce}.
In Ref.\cite{Ce}, by introducing noncommutativity on the space sheet of
the wrapped supermembrane, a consistent truncation of the space-sheet
degrees of freedom was proposed, where it was pointed out that the
underlying mathematical structure is an affine Lie algebra.

The purpose of this paper is to give a concrete matrix representation
of the infinite dimensional Lie algebra in Ref.\cite{Ce} 
and obtain the matrix string
theory having the boundary conditions for the matrix variables
corresponding to the wrapped supermembrane.
Note that the boundary conditions were assumed in Ref.\cite{SY}
but they are derived here.
Since this method relies neither on Seiberg and Sen's arguments nor
on string dualities, this gives support of the string dualities and
the recovery of eleven dimensional Lorentz invariance in the large-$N$
limit.
Furthermore, we discuss the matrix regularization of the multi-wrapped
supermembrane.

The plan of this paper is as follows.
In the next section, we review the consistent truncation of the
space-sheet degrees of freedom in the wrapped membrane theory and
study the algebraic structure.
In section \ref{S:Rep}, we give a concrete matrix representation for
the algebra.
In section \ref{S:M2M}, we obtain the matrix string theory having the
boundary conditions for the matrix variables corresponding to the
wrapped membrane.
In section \ref{S:MWM}, we embed the configuration of the
multi-wrapped membrane in matrix string theory.
Final section is devoted to conclusion.

%%%%%%%%%%%%%%%%%%%%%%%%%%%%%%%%%%
\section{Consistent truncation for wrapped membrane}\label{S:rev}
%%%%%%%%%%%%%%%%%%%%%%%%%%%%%%%%%%
It is well known that Matrix theory can be obtained by truncating the
infinite space-sheet degrees of freedom in the light-cone
supermembrane action on $R^{11}$ to the finite ones.
On the other hand, as to the light-cone wrapped supermembrane on
$R^{10}\times S^1$, the truncation to finite degrees of freedom fails
\cite{SY,Ce}. In particular,
it was pointed out that in the wrapped supermembrane action,
the consistent truncation is for the target-space coordinates to take
values in the representation of an affine Lie algebra \cite{Ce}.
In this section, we review the discussion in Ref.\cite{Ce}.

We can truncate the degrees of freedom of the space-sheet coordinates
$(\sigma,\rho)$ by introducing the noncommutativity
$[\,\sigma,\rho\,]=i\mtheta$ ($\mtheta$: constant).
This noncommutativity is encoded in the star product of functions on
the space sheet,
\begin{equation}
 f*g=f\exp\left(i\frac{1}{2}\mtheta\,\epsilon^{\alpha\beta}\,
    \overleftarrow{\p}_{\alpha}\overrightarrow{\p}_{\beta}\right)\,g.
	\qquad(\alpha,\beta=\sigma,\rho)
\end{equation}
Then, the star-commutator for Fourier modes on the space sheet
is
given by\footnote{For simplicity, we consider only toroidal membrane
in this paper. Recently, the space-sheet topology in the matrix
regularized membrane was discussed in Ref.\cite{Shi}.} \cite{FZ}
\begin{equation}
  \scom{e^{ik_1\sigma+ik_2\rho},e^{ik'_1\sigma+ik'_2\rho}}
   =-2i\sin\left(\frac{1}{2}\mtheta\,k\times k'\right)\,
	e^{i(k_1+k'_1)\sigma+i(k_2+k'_2)\rho}.
\label{alg}
\end{equation}
In the $\mtheta\to0$ limit, the space-sheet Poisson bracket is
obtained,
\begin{equation}
 \{\,f,g\,\}=-i \lim_{\mtheta\to 0}\mtheta^{-1}\,\scom{f,g}\,.
\end{equation}
Henceforth, we set $\mtheta=4\pi/N$ ($N=2M+1$ : odd number).
Then, the Fourier modes $e^{ipN\sigma}$, $e^{irN\rho}$
($p,r\in\mathbb{Z}$) commute with any modes and hence they are central
elements in the star-commutator algebra.
This means that they can be consistently modded out from the
star-commutator algebra, since left and right multiplications coincide
on any modes. Thus we can identify them with the identity operator
and obtain the following equivalence relation,
\begin{eqnarray}
 e^{i(k_1+pN)\sigma+ik_2\rho}&\approx&
	 e^{ik_1\sigma+ik_2\rho},\label{ide1}\\
 e^{ik_1\sigma+i(k_2+rN)\rho}&\approx&
	 e^{ik_1\sigma+ik_2\rho}.\label{ide2}
\end{eqnarray}
Under the identification, we can truncate the infinite dimensional
algebra to the finite dimensional algebra $\mathfrak{u}(N)$
consistently. Then, the mode numbers of $e^{ik_1\sigma+ik_2\rho}$ are
restricted to $k_1,k_2=0,\pm1,\pm2,\cdots,\pm M$.
If we adopt such a consistent truncation for the light-cone
supermembrane on $R^{11}$, we can obtain Matrix theory.

In the case of the wrapped membrane, we need to add a linear function
$\rho$ representing the wrapping to the generators of the
star-commutator algebra.
Then the star commutators are given by eq.(\ref{alg}) and
\begin{equation}
 \scom{\rho,e^{ik_1\sigma+ik_2\rho}}=
	\frac{4\pi k_1}{N}\, e^{ik_1\sigma+ik_2\rho}.\label{alg2}
\end{equation}
Thus, in this case,  we cannot truncate this star-commutator algebra
to a finite dimensional one because the star commutator
$\scom{\rho,e^{ipN\sigma}}=4\pi p e^{ipN\sigma}$ indicates that
$e^{ip N\sigma}$ cannot be the central elements and hence the
equivalence (\ref{ide1}) is not valid.
On the other hand,  $e^{iqN\rho}$ are the central elements and the
equivalence (\ref{ide2}) is still valid.
Then, we can truncate only the Fourier modes with respect to $\rho$
and the truncated generators are given by
$\{e^{ik_1\sigma+ik_2\rho},\rho\,|\,k_1=0,\pm1,\pm2,\cdots,\pm\infty,
\,k_2=0,\pm1,\pm2,\cdots,\pm M\}$ \cite{Ce}.
Note that although we cannot identify $e^{ipN\sigma}$ with the central
elements, they form an ideal of the truncated star-commutator algebra.
Hence this algebra is not simple and henceforth we restrict to the
quotient by this ideal in this section.
In the next section, we will comment on the ideal.

Although this quotient is infinite dimensional,
the rank is finite. Actually, we can adopt $N$ generators
$\{e^{ik\rho},\rho\,|\,k=\pm 1,\pm 2,\cdots,\pm M\}$
as the Cartan subalgebra generators.
We take the basis of the Cartan subalgebra generators as follows,
\begin{eqnarray}
 H^{k}&=&\frac{1}{N}\sum_{l=-M,\,l\ne 0}^{M}\lambda^{kl}
	(\lambda^{-l}-\lambda^{l})\,e^{il\rho},\quad
	(k=0,\pm 1,\pm 2,\cdots,\pm M)\label{Car.1}\\
 D&=&\frac{1}{4\pi}\,\rho-\frac{1}{N}\sum_{l=-M,\,l\ne 0}^{M}
	\frac{1}{\lambda^{-l}-\lambda^{l}}\,e^{il\rho},\label{Car.2}
\end{eqnarray}
where $\lambda \equiv e^{2\pi i/N}$. Note that $\lambda$ has the
following property,
\begin{eqnarray}
 \sum_{l=-M}^{M}\lambda^{kl}&=&N\delta_{k,0}^{(N)},\label{lambda}
\end{eqnarray}
where the indices of the Kronecker symbol $\delta_{k,l}^{(N)}$ are
understood to be modulo $N$.
In eq.(\ref{Car.1}), the index $k$ runs from $-M$ to $M$.
Thus, at first sight the number of the generators in eq.(\ref{Car.1})
seems to be $N\,(=2M+1)$.
However, the number of the independent generators is $N-1$.
Actually, $H^0=-\sum_{l=-M,\,l\ne 0}^{M}H^l$ due to eq.(\ref{lambda})
and hence $H^0$ is not independent.
By using eq.(\ref{lambda}), eqs.(\ref{Car.1}) and (\ref{Car.2}) are
rewritten by
\begin{eqnarray}
 e^{ik\rho} &=&\frac{1}{\lambda^{-k}-\lambda^{k}}
	\sum_{l=-M}^{M}\lambda^{-kl}H^{l},\quad
	(k=\pm 1,\pm 2,\cdots,\pm M)\label{Car.1'}\\
 \rho&=&4\pi D+\frac{4\pi}{N}\sum_{k=-M}^{M}\left\{
	\sum_{l=-M,\,l\ne 0}^{M}
	\frac{\lambda^{-kl}}{(\lambda^{-l}-\lambda^{l})^2}
	\right\}\,H^k\label{Car.2'}.
\end{eqnarray}
As for the remaining infinite raising and lowering generators, we take
the following basis,
\begin{eqnarray}
 E^{k}_{pN+q}&=&\frac{1}{N}\sum_{l=-M}^{M}\lambda^{kl}\,
	e^{i(pN+q)\sigma+il\rho},\hspace{12ex}(p=0,\pm1,\pm2,\cdots)
	\label{others1}\\
 E^{k}_{pN}&=&\frac{1}{N}\sum_{l=-M,\,l\ne 0}^{M}\lambda^{kl}
    (\lambda^{-l}-\lambda^{l})\,e^{ipN\sigma+il\rho},
	\quad(p=\pm1,\pm2,\cdots)\label{others2}
\end{eqnarray}
where $k=0,\pm1,\cdots,\pm M$, $q=\pm1,\pm2,\cdots,\pm M$.
Note that the following relations hold,
\begin{equation}
  E^0_{pN}=-\sum_{l=-M,\, l\ne0}^{M}E_{pN}^l,
  \quad (E^k_{pN})^\dagger=E^{k}_{-pN},
  \quad (E^k_{pN+q})^\dagger=E^{k}_{-(pN+q)}\,.
\end{equation}
Eqs.(\ref{others1}) and (\ref{others2}) are inverted as follows,
\begin{eqnarray}
  e^{i(pN+q)\sigma+ik\rho}&=&
	\sum_{l=-M}^{M}\lambda^{-kl}E_{pN+q}^{l}\,,\hspace{9ex}
	(k=0,\pm 1,\pm 2,\cdots,\pm M)\label{others1'}\\
  e^{ipN\sigma+ik\rho} &=&\frac{1}{\lambda^{-k}-\lambda^{k}}
	\sum_{l=-M}^{M}\lambda^{-kl}E_{pN}^{l}\,.\quad
	(k=\pm 1,\pm 2,\cdots,\pm M)\label{others2'}
\end{eqnarray}

\begin{figure}[htbp]
\centerline{\epsfxsize=102mm\epsfbox{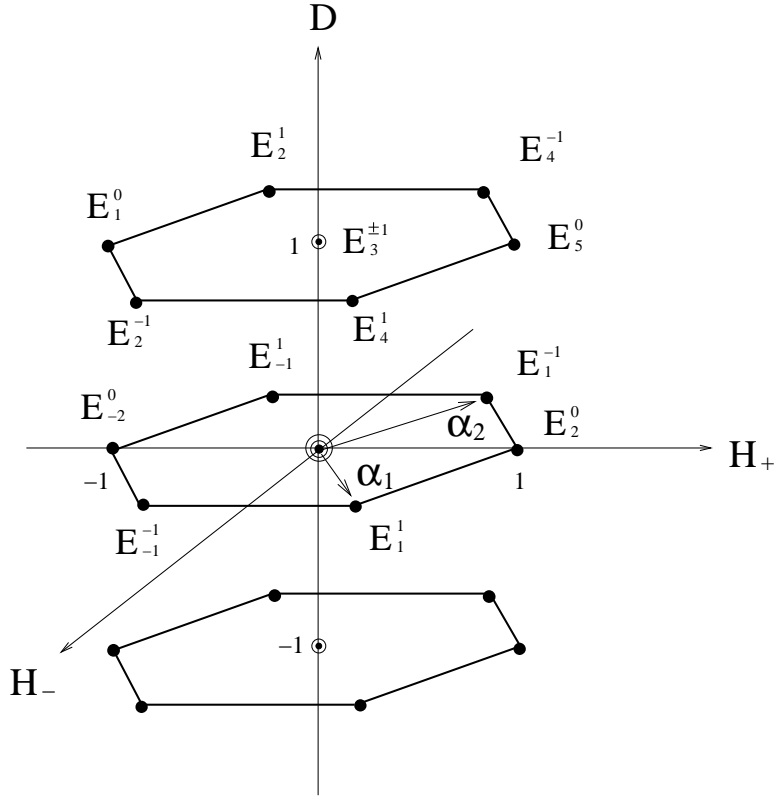}}
\caption{The root system in the $N=3$ case. Here, $\{H_{\pm},D\}$ are
the Cartan subalgebra generators. $\alpha_{1}, \alpha_{2}$ are simple
roots of the zero-mode subalgebra.}\label{fig1}
\end{figure}

{}From eqs.(\ref{alg}) and (\ref{alg2}), we obtain the following
commutators for the generators (\ref{Car.1}), (\ref{Car.2}),
(\ref{others1}) and (\ref{others2}), although the calculation is a bit
lengthly,
\begin{eqnarray}
 \scom{H^{k},H^l} &=&0,\label{com.-1}\\
 \scom{H^{k},D}&=&0,\label{com.0}\\
 \scom{H^{k},E_{pN}^l}&=&0,\label{com.1}\\
 \scom{H^{k},E_{pN+q}^l}&=&(\delta_{k-l+q-1,0}^{(N)}
	-\delta_{k-l+q+1,0}^{(N)}-\delta_{k-l-q-1,0}^{(N)}
	+\delta_{k-l-q+1,0}^{(N)})\,E_{pN+q}^l,\label{com.2}\\
 \scom{D,E_{pN}^l}&=&pE_{pN}^l,\label{com.3}\\
 \scom{D,E_{pN+q}^l}&=&\omega(l,pN+q)\,E_{pN+q}^l\,,\nn
  &&\qquad \biggl(\omega(l,pN+q)\equiv
	 p+\mbox{sgn}(q)\sum_{s=0}^{|q|-1}
	\delta^{(N)}_{2s,\,l+|q|-1}\biggr)\label{com.4}\\
 \scom{E_{pN}^{k},E_{rN}^l}&=&0,\label{com.5}\\
 \scom{E_{pN}^{k},E_{rN+s}^l}&=&(\delta_{k-l+s-1,0}^{(N)}
	-\delta_{k-l+s+1,0}^{(N)}-\delta_{k-l-s-1,0}^{(N)}
	+\delta_{k-l-s+1,0}^{(N)})\,E_{(p+r)N+s}^l,\label{com.6}\\
 \scom{E_{pN+q}^{k},E_{rN+s}^l}&=&\left\{\begin{array}{ll}
   \DS\mbox{sgn}(q)\sum^{|q|-1}_{t=0}H^{2t+k-|q|+1}\,,
	&(k=l,q+s=0,p+r=0)\\[20pt]
   \DS\mbox{sgn}(q)\sum^{|q|-1}_{t=0}E_{(p+r)N}^{2t+k-|q|+1}\,,
	&(k=l,q+s=0,p+r\ne0)\\[20pt]
   E^{k+s}_{(p+r)N+q+s}\,,&
      \lower8pt\hbox{$\left(\begin{array}{c}
	k-l+q+s=0\quad\mbox{mod $N$}\\
        k-l-q-s\ne0\quad\mbox{mod $N$}\end{array}\right)$}\\[22pt]
   -E^{k-s}_{(p+r)N+q+s}\,,
      & \lower8pt\hbox{$\left(\begin{array}{c}
	k-l+q+s\ne0\quad\mbox{mod $N$}\\
	k-l-q-s=0\quad\mbox{mod $N$}\end{array}\right)$}\\[20pt]
    0\,.&~\mbox{otherwise}\label{com.7}\end{array}\right.
\end{eqnarray}
{}From these star commutators, we can obtain the root system of the
quotient.
For simplicity, we consider the $N=3~(M=1)$ case, first.
The root system is given in figure\ref{fig1}, where
we have changed the basis of the Cartan subalgebra generators
$\{H^k,D|k=\pm1\}\to\{H_{\pm},D\}\ (H^{\pm 1}=H_+ \pm
\sqrt{3}H_-)$.\footnote{$H^{k}\ (k=\pm1)$ are the Cartan subalgebra
generators in the Chevalley basis of $\mathfrak{su}(3)$, while
$H_{\pm}$ are those in the ordinary Cartan-Weyl basis of
$\mathfrak{su}(3)$.}
In figure\ref{fig1}, we see the infinite series of the subalgebra
$\mathfrak{su}(3)$ in the direction of $D$, where $D$ is a derivation.
%%%%%%%%%%%%%%%%%%%%%%%%%%
%%%%%%%%%%%%%%%%%%%%%%%%%%
In this root system, the nine generators commute with $D$:
One of them is $D$ itself and the remaining eight generators are
$H^{\pm1}$, $E^{\pm1}_{\pm1}$ and $E^0_{\pm2}$ which constitute the
zero-mode subalgebra of the loop algebra over $\mathfrak{su}(3)$.
Thus this root system agrees with that of affine $\mathfrak{su}(3)$
{\sl except for a central element}.\footnote{With a finite
dimensional simple Lie algebra $\bar{\mathfrak{g}}$, the (untwisted)
affine Lie algebra $\mathfrak{g}$ is obtained by extending
$\bar{\mathfrak{g}}$ to the infinite dimensional Lie algebra
$\bar{\mathfrak{g}}_{\mbox{\tiny loop}}$ of the loop algebra,
centrally extending $\bar{\mathfrak{g}}_{\mbox{\tiny loop}}$ and
adding a derivation $D$, i.e.,
$\mathfrak{g}=\bar{\mathfrak{g}}_{\mbox{\tiny loop}}\oplus
\mathbb{C}K\oplus\mathbb{C}D$, where $K$ is a central element (See
e.g., Ref.\cite{Fuc}). Thus, in order for this root system to agree
with that of an affine Lie algebra, we need to centrally extend the
star commutators (\ref{alg}) and (\ref{alg2}) (or
(\ref{com.-1})-(\ref{com.7})). The central extension would be related
to anomaly in the supermembrane.}
In the zero-mode subalgebra, root vectors $\alpha_1,\alpha_2$
corresponding to the generators $E^{\pm 1}_1$ are simple roots.

The analysis of the general $N$ case is performed similarly.
{}From eqs.(\ref{com.0}), (\ref{com.3}) and (\ref{com.4}), we see
$N^2$ generators which commute with $D$. Among them, $N^2-1$
generators (see a table bellow) constitute the zero-mode subalgebra of
the loop algebra over $\mathfrak{su}(N)$ and the remaining one is $D$
itself.
\[\begin{array}{ll|l}
 \mathrm{generators}&&\mathrm{number}\\\hline
 H^l&(l\ne0)& 2M\\
 E^l_{\pm1}&(l\ne0)& 2M\times2\\
 E^l_{\pm2}&(l\ne\pm1)& (2M-1)\times2\\
 E^l_{\pm3}&(l\ne\pm2,0)& (2M-2)\times2\\
 \quad\vdots &&\quad\vdots\\
 E^l_{\pm (M-1)}&(l\ne\pm(M-2),\pm(M-4),\cdots)&(M+2)\times2\\
 E^l_{\pm M}&(l\ne\pm(M-1),\pm(M-3),\cdots)&(M+1)\times2\\
 E^l_{\pm (M+1)}&(l=\pm(M-1),\pm(M-3),\cdots)&M\times2\\
 E^l_{\pm (M+2)}&(l=\pm(M-2),\pm(M-4),\cdots)&(M-1)\times2\\
 \quad\vdots &&\quad\vdots\\
 E^l_{\pm(N-2)}&(l=\pm1)& 2\times2\\
 E^l_{\pm(N-1)}&(l=0)& 1\times2\\\hline
  &\hfill\mathrm{total}&N^2-1
\end{array}
\]
Thus this root system agrees with that of affine $\mathfrak{su}(N)$
except for a central element.
In the zero-mode subalgebra, $N-1$ root vectors corresponding to the
generators $E_1^{l}\,(l=\pm1,\pm2,\cdots,\pm M)$ are simple roots.
Actually, from the star commutators (\ref{com.-1})-(\ref{com.7}),
$3(N-1)$ generators
$\{E_{\pm1}^{k},H^l\,|\,k,l=\pm1,\pm2,\cdots,\pm M\}$ satisfy the
Chevalley-Serre relations of $\mathfrak{su}(N)$ (see e.g., Ref.\cite{Fuc}),
\begin{eqnarray}
 \scom{H^{k},H^l}&=&0,\\
 \scom{H^{k},E_{\pm 1}^l}&=&
	\pm (2\delta_{k,l}^{(N)}-\delta_{k-l,-2}^{(N)}
	-\delta_{k-l,2}^{(N)})\,E_{\pm 1}^l=\pm A^{lk}E_{\pm 1}^l,\\
 \scom{E_{+1}^{k},E_{-1}^l}&=&\delta_{k,l}H^k,\\
 (\mbox{ad}_{E^k_{\pm1}})^{1-A^{lk}}(E^l_{\pm1})&=&0,\quad (k\ne l)
\end{eqnarray}
where $A^{lk}$ is the $(l,k)$-component of the Cartan matrix of
$\mathfrak{su}(N)$ and $\mbox{ad}_x (y)\equiv[x,y]$.

%%%%%%%%%%%%%%%%%%%%%%%%%%%%%%%%%%%%%%%%%%%%%%%%%%%%%%%%%%%%%
\section{Representation of the algebra}\label{S:Rep}
%%%%%%%%%%%%%%%%%%%%%%%%%%%%%%%%%%%%%%%%%%%%%%%%%%%%%%%%%%%%%
In this section, we give a concrete representation of the
star-commutator algebra in the previous section.
Actually, we can represent the generators (\ref{Car.1})-(\ref{Car.2})
and (\ref{others1})-(\ref{others2}), which satisfy
eqs.(\ref{com.-1})-(\ref{com.7}), as the $N\times N$ matrices with a
continuous parameter $\theta$,
\begin{eqnarray}
 H^{k} &\to& \left((H^k)_{ab}\right),\qquad (H^k)_{ab}\equiv \delta_{ab}\,
	(\delta^{(N)}_{k,2a}-\delta^{(N)}_{k,2(a-1)}),
	\label{matrix1}\\
 D &\to& \left(\frac{-i}{2}\,\ptheta\,\delta_{ab}\right),
	\label{matrix2}\\
 E_{pN}^{k} &\to& \left((E_{pN}^k)_{ab}\right),\qquad
	(E_{pN}^k)_{ab}\equiv e^{i2p\theta}(H^k)_{ab},\label{matrix3}\\
 E_{pN+q}^k &\to& \left((E_{pN+q}^k)_{ab}\right),\quad
	(E_{pN+q}^k)_{ab}\equiv  e^{i2\omega(k,pN+q)\theta}\,
	\delta^{(N)}_{b-a,q}\,\delta^{(N)}_{a+b,k+1},\label{matrix4}
\end{eqnarray}
where matrix indices $a,b=1,2,\cdots,N$.
It is easy to see that these matrices satisfy the star commutators
(\ref{com.-1})-(\ref{com.7}).
Furthermore, by using the above matrices and
eqs.(\ref{Car.1'})-(\ref{Car.2'}) and
(\ref{others1'})-(\ref{others2'}), the matrix representations of the
linear function $\rho$ and Fourier modes
$e^{ipN\sigma+ik\rho}\ (k=\pm1,\pm2,\cdots,\pm M,\,p\in\mathbb{Z})$,
$e^{i(pN+q)\sigma+ik\rho}\ (k=0,\pm1,\pm2,\cdots,\pm M,\
q=\pm1,\pm2,\cdots,\pm M,\ p\in\mathbb{Z})$ are given by
\begin{eqnarray}
 \rho\!\!\!&\to&\!\!\!-2\pi i\ptheta \left(
    \begin{array}{ccccc}
	1&&&\lower10pt\hbox{\Large 0}&\\[-5pt]
	&1&&&\\ &&1&&\\
	&\lower5pt\hbox{\Large0}&&\ddots&\\&&&&1\end{array}\right)
	+4\pi\left(\begin{array}{ccccc}
	\frac{M}{N}&&&\lower10pt\hbox{\Large 0}&\\[-5pt]
	&\frac{M-1}{N}&&&\\ &&\frac{M-2}{N}&&\\
	&\lower5pt\hbox{\Large0}&&\ddots&\\&&&&-\frac{M}{N}
    \end{array}\right),\label{matrix1'}\\
 e^{ipN\sigma+ik\rho}\!\!\! &\to&\!\!\!
    \tau^p\lambda^{-k}\left(\begin{array}{cccccc}
	1& & & &  \lower10pt\hbox{\Large 0}&\\[-5pt]
	 & \lambda^{-2k}& & & & \\
	 &  & \lambda^{-4k}& & & \\
	 & \lower5pt\hbox{\Large0} & &  & \ddots &\\
	 &  & & & &\lambda^{-2(N-1)k}\end{array}\right),\label{matrix2'}\\
 e^{i(pN+q)\sigma+ik\rho}\!\!\!&\to&\!\!\!\!\!\!\label{matrix3'}
  \left\{\begin{array}{l} \tau^p \lambda^{-k(q+1)}\\
  \times\renewcommand{\arraystretch}{1.2}
	\setlength{\arraycolsep}{3pt}
    \left(\begin{array}{cccccccc}
	\multicolumn{2}{l}{\overbrace{0\hspace{40pt}
	\cdots\hspace{45pt}0}^q} & 1&&&&&\\
	&  & 0 & \lambda^{-2k} &&&& \\
	& &  &   \ddots   & \ddots &&\\
	&  &  &  & & 0 & \lambda^{-2(N-q-1)k}\\
	\lambda^{-2(N-q)k}\tau\hfill\phantom{aaa}&& & &&&0\\
	{}\hspace{10ex}~~\ddots& &&& &  &\vdots\\
	\underbrace{0\hspace{20pt}\cdots\hspace{20pt}0}_{q-1}
	&\lambda^{-2(N-1)k} \tau &0 &&&  &0
    \end{array}\right)\!\!,\, (q>0)\\[8em]
    \tau^{p-1} \lambda^{-k(q+1)}\\
    \times\renewcommand{\arraystretch}{1.2}
	\setlength{\arraycolsep}{3pt}
    \left(\begin{array}{cccccccc}
	\multicolumn{2}{l}{\overbrace{0\hspace{40pt}
	\cdots\hspace{45pt}0}^{N+q}} & 1&&&&&\\
	&  & 0 & \lambda^{-2k}&&&& \\
	& &  &   \ddots   & \ddots& &\\
	&  &  &  & & 0 & \lambda^{-2(-q-1))k}\\
	\lambda^{-2(-q)k}\tau\hfill\phantom{aaa}&& & &&&0\\
	{}\hspace{10ex}~~\ddots& &&& &  &\vdots\\
	\underbrace{0\hspace{20pt}\cdots\hspace{20pt}0}_{N+q-1}
	&\lambda^{-2(N-1)k}\tau &0 &&&  &0
    \end{array}\right)\!\!,\,\,\,(q<0)\end{array}\right.
\end{eqnarray}
where $\tau\equiv e^{2i\theta}$.
So far, we have concentrated on the quotient by the ideal
$\{e^{ipN\sigma}\}\ (p\in{\mathbb Z})$.
However, it is easy to extend the discussion in the previous
section with the ideal included.
Then the matrix representations of the generators are given by
\begin{equation}
 e^{ipN\sigma}\to \tau^p\left(\begin{array}{cccccc}
	1& & & &  \lower10pt\hbox{\Large 0}&\\[-5pt]
	 & 1& & & & \\
	 &  & 1& & & \\
	 & \lower5pt\hbox{\Large0} & &  & \ddots &\\
	 &  & & & &1\end{array}\right).\label{matrix4'}
\end{equation}
We summarize the matrix representations of the Fourier modes
(\ref{matrix2'})-(\ref{matrix4'}) as follows,
\begin{eqnarray}
e^{i(pN+q)\sigma+ik\rho}&\to&\label{matrix5'}
 \left\{\begin{array}{l} \tau^p \lambda^{-k(q+1)}\\
 \times\renewcommand{\arraystretch}{1.2}
	\setlength{\arraycolsep}{3pt}
    \left(\begin{array}{cccccccc}
	\multicolumn{2}{l}{\overbrace{0\hspace{40pt}
	\cdots\hspace{45pt}0}^q} & 1&&&&&\\
	&  & 0 & \lambda^{-2k} &&&& \\
	& &  &   \ddots   & \ddots &&\\
	&  &  &  & & 0 & \lambda^{-2(N-q-1)k}\\
	\lambda^{-2(N-q)k}\tau\hfill\phantom{aaa}&& & &&&0\\
	{}\hspace{10ex}~~\ddots& &&& &  &\vdots\\
	\underbrace{0\hspace{20pt}\cdots\hspace{20pt}0}_{q-1}
	&\lambda^{-2(N-1)k} \tau &0 &&&  &0
    \end{array}\right)\!\!,(q>0)\\[7em]
\tau^p\lambda^{-k}\left(\begin{array}{cccccc}
	1& & & & & \lower10pt\hbox{\Large 0}\\[-5pt]
	 & \lambda^{-2k}& & & & \\
	 &  & \lambda^{-4k}& & & \\
	 & \lower5pt\hbox{\Large0} & &  & \ddots &\\
	 &  & & & &\lambda^{-2(N-1)k}\end{array}\right)
	,\hspace{25pt}(q=0)\\[4em]
\tau^{p-1} \lambda^{-k(q+1)}\\
  \times\renewcommand{\arraystretch}{1.2}
	\setlength{\arraycolsep}{3pt}
    \left(\begin{array}{cccccccc}
	\multicolumn{2}{l}{\overbrace{0\hspace{40pt}
	\cdots\hspace{45pt}0}^{N+q}} & 1&&&&&\\
	&  & 0 & \lambda^{-2k}&&&& \\
	& &  &  \ddots   & \ddots& &\\
	&  &  &  & & 0 & \lambda^{-2(-q-1))k}\\
	\lambda^{-2(-q)k}\tau\hfill\phantom{aaa}&& & &&&0\\
	{}\hspace{10ex}~~\ddots& &&& &  &\vdots\\
	\underbrace{0\hspace{20pt}\cdots\hspace{20pt}0}_{N+q-1}
	&\lambda^{-2(N-1)k}\tau &0 &&&  &0
    \end{array}\right)\!\!,(q<0)\end{array}\right.
\end{eqnarray}
where $k,q=0,\pm1,\pm2,\cdots,\pm M,\,p\in\mathbb{Z}$.

%%%%%%%%%%%%%%%%%%%%%%%%%%%%%%%%%%%%%%%%%%%%%%%%%%%%%%%%%%%%%%%%%%%
\section{From wrapped membrane to matrix string}\label{S:M2M}
%%%%%%%%%%%%%%%%%%%%%%%%%%%%%%%%%%%%%%%%%%%%%%%%%%%%%%%%%%%%%%%%%%%
In this section, we show that the consistent truncation of the
light-cone wrapped supermembrane on $R^{10}\times S^1$ leads to matrix
string theory.
In particular, by using the matrix representations in the previous
section, we can derive the boundary conditions of the matrix variables
corresponding to the wrapped supermembrane.

Our starting point is the action of the light-cone wrapped
supermembrane on $R^{10}\times S^1$,\footnote{Precisely speaking,
when the membrane has the non-trivial space-sheet topology, we need to
impose the global constraints to the action (\ref{MLCgauge})
\cite{UY2}. However, for simplicity, such constraints are ignored in
this paper.}
(Here we just write it only with the bosonic degrees of
freedom. Fermions are straightforwardly included.)
\begin{eqnarray}
  S_{WM}&=&\frac{LT}{2}\int d\tau \int_0^{2\pi}d\sigma d\rho
    \left[(D_{\tau}X^i)^2 -\frac{1}{2L^2}\{X^{i},X^{j}\}^2\right]
	\label{MLCgauge},\\
  &&D_{\tau}X^i=\ptau X^i -\frac{1}{L}\{A,X^i\},
\end{eqnarray}
where $i,j=k,9\,(k=1,\cdots,8)$ and $L$ is a radius of
the target space $S^1$.
We take $X^9$ as the $S^1$ direction in the action and
$X^i, A$ are Fourier expanded as
\begin{eqnarray}
 X^9&=&wL\rho + Y\nn
    &=&wL\rho +\sum_{k_1,k_2=-\infty}^{\infty}
	 Y_{(k_1,k_2)}\,e^{ik_1\sigma+ik_2\rho},\\
 X^k&=&\sum_{k_1,k_2 = -\infty}^{\infty}X^k_{(k_1,k_2)}\,
	e^{ik_1\sigma+ik_2\rho},\\
 A&=&\sum_{k_1,k_2=-\infty}^{\infty}A_{(k_1,k_2)}\,
	e^{ik_1\sigma+ik_2\rho},
\end{eqnarray}
where $w(\ne 0)$ is a wrapping number.
Now we introduce the noncommutativity $[\sigma,\rho]=4\pi i/N$
on the space-sheet and carry out the consistent truncation,
\begin{eqnarray}
 X^9&=&wL\rho +\sum_{k_1=-\infty}^{\infty}\sum_{k_2=-M}^{M}
	Y_{(k_1,k_2)}\,e^{ik_1\sigma+ik_2\rho}\nn
    &=&wL\rho+ \sum_{p=-\infty}^{\infty}\sum_{q=-M}^{M}\sum_{k=-M}^{M}
	Y_{(pN+q,k)}\,e^{i(pN+q)\sigma+ik\rho},\label{mode1}\\
 X^k&=&\sum_{k_1=-\infty}^{\infty}\sum_{k_2=-M}^{M}X^k_{(k_1,k_2)}\,
	e^{ik_1\sigma+ik_2\rho}=\sum_{p=-\infty}^{\infty}
	\sum_{q=-M}^{M}\sum_{k=-M}^{M}
	X^k_{(pN+q,k)}\,e^{i(pN+q)\sigma+ik\rho},\label{mode2}\\
 A&=&\sum_{k_1=-\infty}^{\infty}\sum_{k_2=-M}^{M}A_{(k_1,k_2)}\,
	e^{ik_1\sigma+ik_2\rho}=\sum_{p=-\infty}^{\infty}
	\sum_{q=-M}^{M}\sum_{k=-M}^{M} A_{(pN+q,k)}\,
	e^{i(pN+q)\sigma+ik\rho}.\label{mode3}
\end{eqnarray}
By using eq.(\ref{matrix5'}),
the truncated Fourier modes are represented by  $N\times N$ matrices
with a continuous parameter $\theta$ ($X=Y,X^k,A$),
\begin{eqnarray}
 X&=&\sum_{p=-\infty}^{\infty}\sum_{q=-M}^{M}\sum_{k=-M}^{M}
	X_{(pN+q,k)}\,e^{i(pN+q)\sigma+ik\rho}\nn
  &\to& X(\theta)=\sum_{p=-\infty}^{\infty}\sum_{q=-M}^{M}
	\sum_{k=-M}^{M} X_{(pN+q,k)}\,\tau^p\,
	\lambda^{-k(q+1)}M^k_q,\label{X_theta}
\end{eqnarray}
where
\begin{eqnarray}
 M^k_q=\left\{\begin{array}{l}
    \renewcommand{\arraystretch}{1.2}\setlength{\arraycolsep}{3pt}
    \left(\begin{array}{cccccccc}
	\multicolumn{2}{l}{\overbrace{0\hspace{40pt}
	\cdots\hspace{45pt}0}^q} & 1&&&&&\\
	&  & 0 & \lambda^{-2k} &&&& \\
	& &  &  \ddots   & \ddots &&\\
	&  &  &  & & 0 & \lambda^{-2(N-q-1)k}\\
	\lambda^{-2(N-q)k}\tau\hfill\phantom{aaa}&& & &&&0\\
	{}\hspace{10ex}~~\ddots& &&& &  &\vdots\\
	\underbrace{0\hspace{20pt}\cdots\hspace{20pt}0}_{q-1}
	&\lambda^{-2(N-1)k} \tau &0 &&&  &0
    \end{array}\right),\hspace{30pt}(q>0)\\\\
  \left(\begin{array}{cccccc}
	1& & & & & \lower10pt\hbox{\Large 0}\\[-5pt]
	 & \lambda^{-2k}& & & & \\
	 &  & \lambda^{-4k}& & & \\
	 & \hbox{\Large0} & &  & \ddots &\\
	 &  & & & &\lambda^{-2(N-1)k}\end{array}\right)\!,
	\hspace{120pt}(q=0)\\[4em]
  \renewcommand{\arraystretch}{1.2}\setlength{\arraycolsep}{3pt}
   \left(\begin{array}{cccccccc}
	\multicolumn{2}{l}{\overbrace{0\hspace{40pt}
	\cdots\hspace{45pt}0}^{N+q}} & \tau^{-1}&&&&&\\
	&  & 0 & \lambda^{-2k}\tau^{-1}&&&& \\
	& &  &  \ddots   & \ddots &&\\
	&  &  &  & & 0 & \lambda^{-2(-q-1))k}\tau^{-1}\\
	\lambda^{-2(-q)k}\hfill\phantom{aaa}&& & &&&0\\
	{}\hspace{10ex}~~\ddots& &&& &  &\vdots\\
	\underbrace{0\hspace{20pt}\cdots\hspace{20pt}0}_{N+q-1}
	&\lambda^{-2(N-1)k}&0 &&&  &0
    \end{array}\right)\!\!.(q<0)\end{array}\right.
\end{eqnarray}
As an example, we give the matrix representation for the case of
$N=3$ explicitly,
\begin{eqnarray}
X^{(N=3)}(\theta)=\sum_{p=-\infty}^{\infty}\sum_{k=-1}^{1}
  \tau^p\left(\renewcommand{\arraystretch}{1.5}
    \begin{array}{cccc}
  X_{(3p,k)}\lambda^{-k}&X_{(3p+1,k)}\lambda^{-2k}
	&\tau^{-1}X_{(3p-1,k)}\\
  X_{(3p-1,k)}\lambda^{-2k}&X_{(3p,k)}\lambda^{-3k}
	& X_{(3p+1,k)}\lambda^{-4k}\\
  \tau X_{(3p+1,k)}&X_{(3p-1,k)}\lambda^{-4k}
	&X_{(3p,k)}\lambda^{-5k}\end{array}\right).
\end{eqnarray}

We consider the double-dimensional reduction from the wrapped
supermembrane on $R^{10}\times S^1$ to type-IIA superstring on
$R^{10}$.
Classically, this is to remove the non-zero Fourier modes
with respect to $\rho$ by hand \cite{DHIS}.
After such a reduction, the matrix representation $X(\theta)|_{DDR}$
is given by
\begin{eqnarray}
  X(\theta)|_{DDR}=\sum_{p=-\infty}^{\infty}\sum_{q=-M}^{M}
	X_{(pN+q,0)}\tau^pM_q^{0}\,.\label{DDR}
\end{eqnarray}
In this case also, we just give the matrix representation in the $N=3$
case,
\begin{eqnarray}
  X^{(N=3)}(\theta)|_{DDR}=\sum_{p=-\infty}^{\infty}\tau^p
    \left(\renewcommand{\arraystretch}{1.3}\begin{array}{cccc}
	X_{(3p,0)}&X_{(3p+1,0)}&\tau^{-1}X_{(3p-1,0)}\\
	X_{(3p-1,0)}&X_{(3p,0)}& X_{(3p+1,0)}\\
	\tau X_{(3p+1,0)}&X_{(3p-1,0)}&X_{(3p,0)}
    \end{array}\right).
\end{eqnarray}
Note that $X(\theta)|_{DDR}$ is {\sl not} a diagonal matrix, even
though only the zero-modes w.r.t.\ $\rho$ have been extracted in
$X(\theta)|_{DDR}$.
This matrix is represented in the basis where the zero modes
w.r.t.\ $\sigma$  are placed diagonally.
{}From the physical point of view, however, since the zero modes
w.r.t.\ $\rho$ are identified with the coordinates of type-IIA
superstring, the basis where the zero modes with respect to $\rho$ are
placed diagonally seems to be natural.
Actually, in Ref.\cite{SY}, in the latter basis, a correspondence of
the wrapped supermembrane with matrix string was discussed.
Hence we diagonalize the matrix (\ref{DDR}).
Actually, we can diagonalize it with the following unitary matrix $P$,
\begin{eqnarray}
 P&=&TS,\label{P}\\
  T &\equiv& \left(\begin{array}{cccccc}
	\tau^{-\frac{M}{N}}& & & & & \lower10pt\hbox{\Large 0}\\[-5pt]
	 & \tau^{-\frac{(M-1)}{N}}& & & & \\
	 &  & \tau^{-\frac{(M-2)}{N}}& & & \\
	 & \hbox{\Large0} & &  & \ddots &\\
	 &  & & & &\tau^{\frac{M}{N}}\end{array}\right),\\
 S&\equiv& \frac{1}{\sqrt{N}}
    \renewcommand{\arraystretch}{1.5}\left(\begin{array}{ccccc}
	1& 1& 1&  & 1\\
	1& \lambda^2& \lambda^4& \cdots &\lambda^{2(N-1)}\\
	1 & \lambda^4 & \lambda^8& \cdots &\lambda^{4(N-1)} \\
	\vdots &\vdots  &\vdots &  \ddots &\vdots\\
	1&\lambda^{2(N-1)}&\lambda^{4(N-1)}&\cdots&\lambda^{2(N-1)^2}
    \end{array}\right).
\end{eqnarray}
Then we have
\begin{eqnarray}
P^{\dagger}X(\theta)|_{DDR}P&=&S^{\dagger}T^{\dagger}\,
	X(\theta)|_{DDR}\,TS\nn
  &=&S^{\dagger}\,\left(\sum_{p=-\infty}^{\infty}\sum_{q=-M}^{M}
	X_{(pN+q,0)}\,\tau^{p+\frac{q}{N}}\,V^q\right)\,S\nn
  &=&\sum_{p=-\infty}^{\infty}\sum_{q=-M}^{M}X_{(pN+q,0)}
	\tau^{p+\frac{q}{N}}U^q\nn
  &=&\sum_{p=-\infty}^{\infty}\sum_{q=-M}^{M}X_{(pN+q,0)}\,
	\tau^{p+\frac{q}{N}}\left(\begin{array}{cccccc}
	 1& & & & & \lower10pt\hbox{\Large 0}\\[-5pt]
	 & \lambda^{2q}& & & & \\
	 &  & \lambda^{4q}& & & \\
	 &  & &  \lambda^{6q}&& \\
	 & \hbox{\Large0} & &  & \ddots &\\
	 &  & & & &\lambda^{2(N-1)q}\end{array}\right)\nn
 &\equiv& \left(\begin{array}{cccccc}
	 x_1(\theta)& & & & & \lower10pt\hbox{\Large 0}\\[-5pt]
	 & x_2(\theta)& & & & \\
	 &  & x_3(\theta)& & & \\
	 &  & &  x_4(\theta)&& \\
	 & \hbox{\Large0} & &  & \ddots &\\
	 &  & & & &x_N(\theta)\end{array}\right),\label{eq:DDRdig}
\end{eqnarray}
where $U$ and $V$ are the clock and shift matrices, respectively,
\begin{eqnarray}
  U &=& \left(\begin{array}{cccccc}
	1& & & & & \lower10pt\hbox{\Large 0}\\[-5pt]
	 & \lambda^2& & & & \\
	 &  & \lambda^4& & & \\
	 &  & &  \lambda^6 && \\
	 & \hbox{\Large0} & &  & \ddots &\\
	 &  & & & &\lambda^{2(N-1)}\end{array}\right),\\
 V &=& \left(\begin{array}{ccc@{}c@{}cc}
	0& 1&  & & & \\
	 & 0& 1& & & \\
	\vdots &  &  &\ddots&\ddots  & \\
	0&  &  & & 0& 1\\
	1& 0&  &\cdots & &0\end{array}\right) .
\end{eqnarray}
$U$ and $V$ satisfy $U^N=V^N=1$. For $q<0$,
$U^q\equiv(U^{\dagger})^{-q},\,V^q\equiv(V^{\dagger})^{-q}$
and $S^{\dagger}VS=U,\,S^{\dagger}US=V^{-1}$.
The diagonal elements $x_a(\theta)$ (\ref{eq:DDRdig}) in matrix string
theory are expressed by the Fourier coefficients in the wrapped
supermembrane theory,
\begin{eqnarray}
 x_a(\theta)&=&\sum_{p=-\infty}^{\infty}\sum_{q=-M}^{M}
	X_{(pN+q,0)}\,e^{2i(p+\frac{q}{N})\theta}\,
	e^{i\frac{4(a-1)\pi}{N}q}\nn
  &=&\sum_{n=-\infty}^{\infty}X_{(n,0)}\,
	e^{2i\frac{n}{N}(\theta+2(a-1)\pi)}.\label{x}
\end{eqnarray}
Then it is easy to see that these diagonal elements satisfy the
following boundary conditions,
\begin{eqnarray}
 x_a(\theta+2\pi)&=&x_{a+1}(\theta),\quad
	(a=1,\cdots,N-1)\label{bcD1}\\
 x_{N}(\theta+2\pi)&=&x_{1}(\theta).\label{bcD2}
\end{eqnarray}
Thus we have derived that via the double-dimensional reduction,
the wrapped supermembrane corresponds to a long string, which is
given by the boundary conditions (\ref{bcD1})-(\ref{bcD2}), in matrix
string theory.
 
Next, we consider the $k$-th Fourier mode ($k>0$) with respect to
$\rho$,
\begin{eqnarray}
  X(\theta)|_{\mbox{$k$-th}}&=&\sum_{p=-\infty}^{\infty}
	\sum_{q=-M}^{M}X_{(pN+q,k)}\,\tau^p\,\lambda^{-k(q+1)}M^k_q.
\end{eqnarray}
By using the same unitary matrix $P$ (\ref{P}), we represent this
matrix in the basis where the zero modes with respect to $\rho$ become
the diagonal elements,
\begin{eqnarray}
 P^{\dagger}X(\theta)|_{\mbox{\small $k$-th}}P
  &=&S^{\dagger}T^{\dagger}X(\theta)|_{\mbox{\small $k$-th}}\,TS\nn
  &=&S^{\dagger}\left(\sum_{p=-\infty}^{\infty}\sum_{q=-M}^{M}
	X_{(pN+q,k)}\,\tau^{p+\frac{q}{N}}\,
	\lambda^{-k(q+1)}U^{-k}V^q\right)\,S\nn
 &=&\sum_{p=-\infty}^{\infty}\sum_{q=-M}^{M}X_{(pN+q,k)}
	\,\tau^{p+\frac{q}{N}}\,\lambda^{-k(q+1)}V^kU^q\nn
 &\equiv&\!\!\!\!\left(\begin{array}{cccccccc}
	\multicolumn{2}{l}{\overbrace{0\hspace{40pt}
	\cdots\hspace{45pt}0}^k} & X_{1\,k+1}(\theta)&&&&&\\
	&  & 0 &  X_{2\,k+2}(\theta) &&&& \\
	& &   & \ddots   & \ddots&&\\
	&  &  &  & 0& & X_{N-k\,N}(\theta)\\
	X_{N-k+1\,1}(\theta)\hfill\phantom{aaa}&& & &&&0\\
	{}\hspace{10ex}~~\ddots& &&& &  &\vdots\\
	\underbrace{0\hspace{20pt}\cdots\hspace{20pt}0}_{k-1}
	&X_{N\,k}(\theta) &0 &&&  &0
    \end{array}\right).\nonumber\\\label{k}
\end{eqnarray}
The non-zero matrix elements,
\begin{eqnarray}
 \left.\begin{array}{l} X_{a\,k+a}(\theta)\qquad(a=1,\cdots,N-k)\\
    X_{a\,k+a-N}(\theta)\ \,(a=N-k+1,\cdots,N)\end{array}\right\}
  &=& \sum_{p=-\infty}^{\infty}\sum_{q=-M}^{M}X_{(pN+q,k)}
    e^{i2(p+\frac{q}{N})\theta}e^{i\frac{2k(q-1)\pi }{N}}\,
	e^{i\frac{4(a-1)\pi}{N}q}\nn
  &=&\sum_{n=-\infty}^{\infty}X_{(n,k)}\,
	e^{i2\frac{n}{N}(\theta+2(a-1)\pi)}\,
	e^{i2\pi\frac{n-1}{N}k}\label{X}
\end{eqnarray}
satisfy the following boundary conditions,
\begin{eqnarray}
  X_{a\,k+a}(\theta+2\pi)&=&X_{a+1\,k+a+1}(\theta),\qquad
	(a=1,\cdots,N-k-1)\label{bc1}\\
  X_{N-k\,N}(\theta+2\pi)&=&X_{N-k+1\,1}(\theta),\\
  X_{a\,k+a-N}(\theta+2\pi)&=&X_{a+1\,k+a-N+1}(\theta),\quad
	(a=N-k+1,\cdots,N-1)\\
  X_{N\,k}(\theta+2\pi)&=&X_{1\,k+1}(\theta).\label{bc4}
\end{eqnarray}
In the case of $-k$-th Fourier modes ($k>0$), the matrix is given by
the Hermitian conjugation of eq.(\ref{k}).

Furthermore, by using the unitary matrix $P$ (\ref{P}), the matrix
representation of the linear function $\rho$ (\ref{matrix1'}) is
transformed as follows,
\begin{eqnarray}
P^{\dagger}\rho P&=&S^{\dagger}T^{\dagger}\,\rho\,TS\nn
&=&S^{\dagger}(-2\pi i) \ptheta\left(\begin{array}{ccccc}
	1&&&\lower10pt\hbox{\Large 0}&\\[-5pt]
	&1&&&\\ &&1&&\\
	&\hbox{\Large0}&&\ddots&\\&&&&1\end{array}\right) S\nn
&=&-2\pi i\ptheta\left(\begin{array}{ccccc}
	1&&&\lower10pt\hbox{\Large 0}&\\[-5pt]
	&1&&&\\ &&1&&\\
	&\hbox{\Large0}&&\ddots&\\&&&&1\end{array}\right).
\end{eqnarray}
Thus in the basis where the zero modes w.r.t.\ $\rho$ are
diagonalized, {\sl the matrix representation of the linear function
$\rho$ is proportional to the derivative $-i\p_{\theta}$ times the
unit matrix}.\footnote{Precisely speaking, this statement is not always
correct because the transformation matrix $P$ (\ref{P}) has an
ambiguity of the overall phase $e^{i\alpha\theta}$.
However, even if we have included such a phase factor in eq.(\ref{P}),
the matrix representation of $\rho$ in the transformed basis is
proportional to the unit matrix since the additional term is
proportional to $\alpha$ times the unit matrix. And such an extra term
does not affect matrix string theory (\ref{MS}).}
On the other hand, in the original basis,
the matrix representation (\ref{matrix1'}) is not
proportional to the unit matrix.
Henceforth, all matrices are represented in such basis
as the zero modes w.r.t.\ $\rho$ are diagonalized
and we rewrite $P^{\dagger}X(\theta)P\,\,(X=Y,X^k,A)$ and
$P^{\dagger}\rho P$ to $X(\theta)$ and $\rho$, respectively.
Then, the matrix representations of $X^9, X^k$ and $A$ are given by
\begin{eqnarray}
 X^9&\to&-2\pi i wL \p_{\theta}\left(\begin{array}{ccccc}
	1&&&\lower10pt\hbox{\Large 0}&\\[-5pt]
	&1&&&\\ &&1&&\\
	&\hbox{\Large0}&&\ddots&\\&&&&1\end{array}\right)
 	+Y(\theta),\label{X^9}\\
  X^k&\to&X^k(\theta),\label{X^k}\\
  A&\to&A(\theta).\label{A}
\end{eqnarray}
{}From eqs.(\ref{bcD1})-(\ref{bcD2}) and (\ref{bc1})-(\ref{bc4}),
we find that $Y(\theta), X^k(\theta)$ and $A(\theta)$
satisfy the boundary conditions,
\begin{eqnarray}
  Y(\theta+2\pi)&=&V Y(\theta)V^{\dagger},\label{bc1a}\\
  X^k(\theta+2\pi)&=&V X^k(\theta)V^{\dagger},\\
  A(\theta+2\pi)&=&V A(\theta)V^{\dagger}.\label{bc3a}
\end{eqnarray}
In Ref.\cite{SY}, the boundary conditions were assumed, while they are
derivable in our case.

Finally, we show that after the consistent truncation, the action of
the light-cone wrapped supermembrane on $R^{10}\times S^1$ agrees with
matrix string theory \cite{SY,Ce}.
In such a truncation, the functions $X^9,X^k,A$ of $\sigma$ and $\rho$
are represented by the matrices (\ref{X^9})-(\ref{A}) and
the Poisson bracket and the double integral are represented as follows,
\begin{eqnarray}
 \{\,\cdot\,,\,\cdot\,\} &\to&
	-i\frac{N}{4\pi}[\,\cdot\,,\,\cdot\,],\\
 \int_0^{2\pi}d\sigma d\rho &\to& \frac{2\pi}{N}\int_0^{2\pi}
	d\theta\,\mbox{Tr}.
\end{eqnarray}
{}From these results, the action (\ref{MLCgauge}) in the case of the
single wrapping is mapped to
\begin{eqnarray}
 S_{MS}
   &=&\frac{LT}{2}\int d\tau \frac{\pi N}{2}\int_0^{2\pi}d\theta\,
	\mbox{Tr}\left[ (F_{\tau\theta})^2+
	(D_{\tau}X^k)^2 -(D_{\theta}X^k)^2 \right.\nn
 &&\hspace{5cm}
	\left.+\frac{1}{2(2\pi L)^2}[X^{k},X^{l}]^2\right],\\
  &&F_{\tau\theta}=\frac{2}{N}\ptau Y -\p_{\theta}A
	+i\frac{1}{2\pi L}[A,Y],\\
  &&D_{\tau}X^k=\frac{2}{N}\ptau X^k +i\frac{1}{2\pi L}[A,X^k],\\
 &&D_{\theta}X^k=\p_{\theta} X^k +i\frac{1}{2\pi L}[Y,X^k].
\end{eqnarray}
By rescaling $\tau\to (2/N)\,\tau$, we obtain
\begin{eqnarray}
S_{MS}
 &=&\frac{\pi LT}{2}\int d\tau \int_0^{2\pi}d\theta\,
 \mbox{Tr}\left[ (F_{\tau\theta})^2+
   (D_{\tau}X^k)^2 -(D_{\theta}X^k)^2
+\frac{g^2}{2}[X^{k},X^{l}]^2\right]\!,\label{MS}\\
 &&F_{\tau\theta}=\ptau Y -\p_{\theta}A
+ig[A,Y],\label{MS2}\\
  &&D_{\tau}X^k=\ptau X^k +ig[A,X^k],\label{MS3}\\
&&D_{\theta}X^k=\p_{\theta} X^k +ig[Y,X^k],\label{MS4}
\end{eqnarray}
where $g=1/(2\pi L)$.
The fields $Y(\theta), X^k(\theta)$ and $A(\theta)$
satisfy the boundary conditions (\ref{bc1a})-(\ref{bc3a}).
This action is just a bosonic part of matrix string theory,
i.e., 1+1-dimensional $U(N)$ super Yang-Mills theory.
Thus we have obtained the matrix string theory having the boundary
conditions for the matrix variables corresponding to the wrapped
supermembrane.

%%%%%%%%%%%%%%%%%%%%%%%%%%%%%%%%%%%%%%%%%%%%%%%%%%%%%%%%%%%%%%%%%%%
\section{Multi-wrapped membranes in matrix string theory}\label{S:MWM}
%%%%%%%%%%%%%%%%%%%%%%%%%%%%%%%%%%%%%%%%%%%%%%%%%%%%%%%%%%%%%%%%%%%
In this section, we consider the matrix regularization of the
multi-wrapped supermembrane on $R^{10}\times S^1$ in the light-cone
gauge. Similarly to eqs.(\ref{MS})-(\ref{MS4}), the multi-wrapped
supermembrane action is matrix-regularized as
\begin{eqnarray}
 S_{MS}^{(w)} &=&\frac{\pi LT}{2}\int d\tau \int_0^{2\pi}d\theta\,
     \mbox{Tr}\left[ (F_{\tau\theta})^2+(D_{\tau}X^k)^2
	-(D_{\theta}X^k)^2+\frac{g^2}{2}[X^{k},X^{l}]^2
	\right],\label{MS'}\\
 &&F_{\tau\theta}=\ptau Y -w\p_{\theta}A +ig\,[A,Y],\label{MS2'}\\
 &&D_{\tau}X^k=\ptau X^k +ig\,[A,X^k],\label{MS3'}\\
 &&D_{\theta}X^k=w\p_{\theta} X^k +ig\,[Y,X^k],\label{MS4'}
\end{eqnarray}
where $g=1/(2\pi L)$ and $w$ is the wrapping number.\footnote{In this
section, for simplicity, we consider $w>0$ case only.}
By rescaling $\tau \to \tau/w$, we obtain
\begin{eqnarray}
 S_{MS}^{(w)}&=&\frac{w\pi LT}{2}\int d\tau \int_0^{2\pi}d\theta\,
     \mbox{Tr}\Biggl[ (F^{(w)}_{\tau\theta})^2+(D^{(w)}_{\tau}X^k)^2
	-(D^{(w)}_{\theta}X^k)^2\nn
 &&\hspace{5cm}+\frac{g^{(w)2}}{2}[X^{k},X^{l}]^2
	\Biggr],\label{MS''}\\
 &&F^{(w)}_{\tau\theta}=\ptau Y -\p_{\theta}A
	+ig^{(w)}[A,Y],\label{MS2''}\\
 &&D^{(w)}_{\tau}X^k=\ptau X^k +ig^{(w)}[A,X^k],\label{MS3''}\\
 &&D^{(w)}_{\theta}X^k=\p_{\theta} X^k +ig^{(w)}[Y,X^k],\label{MS4''}
\end{eqnarray}
where $g^{(w)}=1/(2\pi wL)$. In order to see the physical meaning, we
consider the double-dimensional reduction of this action. Classically,
this is to remove the off-diagonal matrix elements by hand.
Then, we obtain the discretized action of a ten-dimensional
superstring with $w$ times the minimal string tension. However, such
objects cannot be incorporated in type-IIA superstring theory. One
possible interpretation is to regard the $w$-wrapped supermembrane as
$w$ fundamental type-IIA superstrings rather than as a ten-dimensional
superstring with $w$ times the minimal string tension \cite{BBS}.
Note that after the double-dimensional reduction, the $w$-dependence
in action (\ref{MS''}) through $g^{(w)}$ is disappeared.

We also can embed the multi-wrapped supermembrane into matrix string
theory, i.e., the single-wrapped supermembrane in the
matrix-regularized form.
By rescaling $\theta\to w\theta$ in eq.(\ref{MS'}), we obtain
\begin{eqnarray}
 S_{MS}^{(w)}&=&w\int d\tau \int_0^{\frac{2\pi}{w}}d\theta\,
	\Lag (Y(w\theta),X^k(w\theta),A(w\theta))\,,\\
 \Lag (Y(\theta),X^k(\theta),A(\theta))
     &=&\frac{\pi LT}{2} \mbox{Tr}\left[ (F_{\tau\theta})^2
	+(D_{\tau}X^k)^2 -(D_{\theta}X^k)^2
	+\frac{g^2}{2}[X^{k},X^{l}]^2\right],\\
 F_{\tau\theta}&=&\ptau Y -\p_{\theta}A+ig[A,Y]\,,\\
 D_{\tau}X^k&=&\ptau X^k +ig[A,X^k]\,,\\
 D_{\theta}X^k&=&\p_{\theta} X^k +ig[Y,X^k]\,.
\end{eqnarray}
Note that the Lagrangian $\Lag (Y(\theta),X^k(\theta),A(\theta))$
is that of matrix string theory, i.e., matrix-regularized Lagrangian
of the single-wrapped supermembrane. This action is rewritten as
follows,
\begin{eqnarray}
 S_{MS}^{(w)}\!\!\!&=&\!\!\!
  \int d\tau \int_0^{\frac{2\pi}{w}}d\theta\,{\Lag}(Y(w\theta),
	X^k(w\theta),A(w\theta))\nn
  &&\hspace{-3ex}+\int d\tau \int_{\frac{2\pi}{w}}^{\frac{4\pi}{w}}
	d\theta\,\Lag (Y(w\theta-2\pi),
	X^k(w\theta-2\pi),A(w\theta-2\pi))\nn
  &&\hspace{1cm}\vdots\nn
  &&\hspace{-3ex}+\int d\tau \int_{\frac{w-1}{w}2\pi}^{2\pi}
	d\theta\,{\Lag}(Y(w\theta-(w-1)2\pi),
	X^k(w\theta-(w-1)2\pi),A(w\theta-(w-1)2\pi))\nn
 \!\!\!&=&\!\!\!
    \int d\tau \int_0^{\frac{2\pi}{w}}d\theta\,{\Lag}(Y(w\theta),
	X^k(w\theta),A(w\theta))\nn
  &&\hspace{-3ex}+\int d\tau \int_{\frac{2\pi}{w}}^{\frac{4\pi}{w}}
	d\theta\,{\Lag}(V^{\dagger}Y(w\theta)V,
	V^{\dagger}X^k(w\theta)A,V^{\dagger}A(w\theta)V)\nn
  &&\hspace{1cm}\vdots\nn
  &&\hspace{-3ex}+\int d\tau \int_{\frac{w-1}{w}2\pi}^{2\pi}
	d\theta\,{\Lag}((V^{w-1})^{\dagger}Y(w\theta)V^{w-1},
	(V^{w-1})^{\dagger}X^k(w\theta)V^{w-1},
	(V^{w-1})^{\dagger}A(w\theta)V^{w-1}),\nonumber
\end{eqnarray}
where we have used the boundary conditions (\ref{bc1a})-(\ref{bc3a}).
Due to gauge invariance, we have
\begin{equation}
 \Lag (Y(\theta),X^k(\theta),A(\theta))=
 \Lag (V^{\dagger}Y(\theta)V,V^{\dagger}X^k(\theta)V,
	V^{\dagger}A(\theta)V).
\end{equation}
Then we obtain
\begin{eqnarray}
 S_{MS}^{(w)}&=&\int d\tau \int_0^{2\pi}d\theta\,
    \Lag (Y(w\theta),X^k(w\theta),A(w\theta))\nn
 &=&\int d\tau \int_0^{2\pi}d\theta\,
    \Lag (Y^{(w)}(\theta),X^{k(w)}(\theta),A^{(w)}(\theta))\,,
\end{eqnarray}
where $X^{(w)}(\theta)\equiv X(w\theta)\  (X=Y,X^k,A)$.
Thus we have succeeded in naturally embedding the multi-wrapped
supermembrane into matrix string theory.

%%%%%%%%%%%%%%%%%%%%%%%%%%%%%%%%%%%%%%%%%%%%%%%%%%%%%%%%%%%%%%%%%%
\section{Conclusion}\label{secCon}
%%%%%%%%%%%%%%%%%%%%%%%%%%%%%%%%%%%%%%%%%%%%%%%%%%%%%%%%%%%%%%%%%%
In this paper, we have given a concrete matrix representation of the
infinite dimensional Lie algebra \cite{Ce} to obtain matrix string
theory via matrix regularization for the wrapped supermembrane on
$R^{10}\times S^1$ in the light-cone gauge.
We have explicitly given the correspondence of
matrix string with the wrapped supermembrane.
That is, in eqs.(\ref{x}) and (\ref{X}), the matrix elements in
matrix string theory are determined completely by the Fourier
coefficients in the wrapped supermembrane theory.
Furthermore, eqs.(\ref{x}) and (\ref{X}) determine the boundary
conditions for the matrix variables in matrix string theory.
We should notice that we have never used the standard Seiberg and
Sen's arguments and string dualities in obtaining the matrix string
theory in this paper. Thus, this method gives support to the string
dualities and the recovery of eleven dimensional Lorentz invariance in
the large-$N$ limit.

\vspace{\baselineskip}

\noindent{\bf Note added:} While finishing the manuscript, a
complementary paper \cite{HI} appeared in the e-print archive, where
matrix string theory is derived with the string dualities and the 9-11
flip. Furthermore, see Ref.\cite{DW} for a different approach to
the wrapped supermembrane.
 
\vspace{\baselineskip}

\noindent{\bf Acknowledgments:}
We would like to thank N. Kitsunezaki for useful discussion.
This work is supported in part by MEXT Grant-in-Aid for
the Scientific Research \#13135212 (S.U.) and JSPS Grant-in-Aid for the
Scientific Research (B)(2) \#14340072 (S.Y.).
%%%%%%%%%%%%%%%%%%%%%%%%%%%%%%%%%

%%%%%%%%%%%%%%%%%%%%%%%%%%%%%%%%%
\end{document}